\documentclass[floatfix,showpacs,amsmath,preprintnumbers,nofootinbib,onecolumn] {revtex4}
\usepackage{graphicx}
\usepackage{colordvi}
\usepackage{amsmath}
\usepackage{amssymb}
\usepackage{latexsym}
    \def\be{\begin{equation}}
    \def\ee{\end{equation}}
    \def\ba{\begin{eqnarray}}
    \def\ea{\end{eqnarray}}
\begin{document}

\title{Brans-Dicke theory and primordial black holes in early matter-dominated era}

\author{B. Nayak and L. P. Singh}
\affiliation{{Department of Physics, Utkal University, Vanivihar,
Bhubaneswar 751004, India}\\
E-mail: {bibeka@iopb.res.in and lambodar\_uu@yahoo.co.in}}

\begin{abstract}
We show that primordial black holes can be formed in the matter-dominated era 
with gravity  
described by the Brans-Dicke theory. Considering an early matter-dominated era
between inflation and reheating, we found that the primordial black holes
 formed during that era evaporate at a quicker rate than those of early radiation-dominated era. Thus, in comparison with latter case, less number of primordial black holes could exist today. Again the constraints on primordial black hole
formation tend towards the larger value than their radiation-dominated era counterparts
 indicating a significant
enhancement in the formation of primordial black holes during the matter-dominaed era.
\end{abstract}
\maketitle
\section{Introduction}
General Theory of Relativity(GTR) \cite{ein}, which was proposed by Einstein in 1916, takes gravitational constant($G$) as a time-independent quantity.
It is a pure tensor theory of gravity. As the extensions of GTR, many scalar tensor theories have been developed by assuming G is a time dependent quantity. 
Among them Brans-Dicke(BD) theory \cite{bdt} is the simplest one. 
In BD theory the gravitational constant is 
set by the inverse of a time-dependent scalar field which couples to gravity 
with a coupling parameter $\omega$. GTR can be recovered from BD theory in 
the limit $\omega \to \infty$ \cite{bam}. BD theory also admits simple 
expanding solutions \cite{mj} for scalar field $\phi(t)$ and scale 
factor $a(t)$ which are compatible with solar system observations \cite{prg}. 
BD theory is also sucessful in explaining many cosmological phenomena such as 
inflation \cite{ls}, early and late time behaviour of the Universe \cite{ss}, 
cosmic acceleration and structure formation \cite{bm}, cosmic acceleration and 
coincidence problem \cite{nb,bn} etc.

It was first predicted by Zeldovich and Novikov \cite{zn} in $1967$ and later by
 Hawking \cite{haw} in $1971$ that black holes could be formed in the early Universe which are known as Primordial Black Holes (PBHs). PBHs may be formed as a 
result of density fluctuation \cite{carr1}, inflation \cite{kmz}, 
phase transition \cite{kp}, bubble collision \cite{kss} and decay of cosmic loops \cite{polzem} etc. 
These black holes are of special interest because their 
masses could be small enough, smallest being of Planck mass $10^{-5}$gm \cite{zn}, to evaporate by the present epoch as a result of 
quantum emission \cite{hawk}. Again PBHs could act as seeds for structure 
formation \cite{mor} and could also form a significant component of 
dark matter \cite{bkp}.

From standard picture of Cosmology, we know that the Universe is 
radiation-dominated just after inflation and it becomes 
matter-dominated much later. So PBHs were expected to be  only formed 
in radiation-dominated era. However, the detailed analysis \cite{kp,ky,pk} 
shows strong enhancement in probability of PBH formation in the 
matter-dominated era compared with the radiation-dominated epoch. 
It has been conjectured by Khlopov et al. \cite{khlopov} and Carr et al. \cite{cgl} 
that there may be an 
early matter-dominated era between the end of the inflation and the onset of 
reheating during which significant PBH formation could occur. It is, 
therefore, an open and interesting problem to investigate PBH formation and 
their evolution in matter-dominated era within the context of 
BD theory, although a number of similar studies have been done in 
radiation-dominated era \cite{bc, nsm, mgs}. 
In this work, we have undertaken such an analysis and show that 
PBHs can indeed be formed in early matter-dominated era. We have also 
studied how it affects various astrophysical constraints through 
the evaporation of PBHs.

\section{PBH in matter-dominated era}
The gravitational field equations for a spatially flat FRW Universe 
with scale factor `$a$' using BD theory are
\be \label{bden}
\frac{\dot{a}^2}{a^2}+\frac{\dot{a}}{a} \frac{\dot{\Phi}}{\Phi}-\frac{\omega}{6} \frac{\dot{\Phi}^2}{\Phi^2}=\frac{8\pi \rho}{3\Phi}
\ee
\ba \label{bpre}
2 \frac{\ddot{a}}{a}+\frac{\dot{a}^2}{a^2}+2 \frac{\dot{a}}{a} \frac{\dot{\Phi}}{\Phi}+\frac{\omega}{2} \frac{\dot{\Phi}^2}{\Phi^2}+\frac{\ddot{\Phi}}{\Phi}=-\frac{8 \pi p}{\Phi}  .
\ea
The wave equation for BD scalar field is
\be \label{bdsc}
\frac{\ddot{\Phi}}{8 \pi}+3 \frac{\dot{a}}{a} \frac{\dot{\Phi}}{8 \pi}=\frac{\rho -3p}{2\omega+3}  .
\ee
Using the above three equations and the perfect fluid equation of state $p=\gamma \rho$, energy conservation equation can be written as
\be \label {econ}
\dot{\rho}+3 \frac{\dot{a}}{a}(\gamma+1) \rho=0  .
\ee
For matter-dominated era $p=0$ which implies $\gamma=0$.
Barrow and Carr \cite{bc} have found that for matter-dominated era, the solutions of above equations are
\ba \label{soln}
\left.
\begin{array}{r}
a(t) \propto t^{\frac{2-n}{3}} \\
G(t) = G_0 (\frac{t_0}{t})^n
\end{array}
\right \}
\ea
where $t_0$ is the present time, $G_0$ is the present value of $G$ and $n=\frac{2}{4+3 \omega}$ . But Solar system 
observations require \cite{bit} $\left|\omega \right| \geq 10^4$. 
Taking $\left|\omega \right| =10^4$, we found $n \approx 0.00007$.\\
Integrating equation (\ref{econ}), one gets
\be \label{vrho}
\rho=\rho_0 (\frac{a}{a_0})^{-3}
\ee
which in conjuction with equation (\ref{soln}) leads to
\be \label{vrhof}
\rho=\rho_0 (\frac{t_0}{t})^{2-n}  .
\ee
We now proceed to discuss about the PBH formation in matter-dominated era. 
Following the analysis of Khlopov \cite{ky}, we assume that the density fluctuation is 
responsible for forming PBHs 
in matter dominated era. This density fluctuation grows to a sufficiently 
homogeneous and isotropic configuration which separates itself from 
cosmological expansion and contracts within its gravitational radius.

Let $t_1$ be the time when contraction starts, 
$r_1$ be the size of the configuration at time $t_1$, 
$S$ be the deviation of configuration from the spherical form at time $t_1$ 
which can be defined as $S=max\{|\gamma_1-\gamma_2|, |\gamma_1-\gamma_3|, |\gamma_2-\gamma_3|\}$, 
where $\gamma_1,~\gamma_2,~\gamma_3$ define the deformation along the three main orthogonal 
axes of the configuration, 
$u$ ( $u \sim \frac{\delta\rho_1}{\rho_1}$) be the inhomogeneity of the density distribution inside the configuration,
and  $\rho_1$ be the mean cosmological density at time $t_1$ .\\
Now equation (\ref{vrhof}) implies,
\be \label {vrhot}
\rho_1=\rho_0 (\frac{t_0}{t_1})^{2-n}  .
\ee
The mean density of the primordial black holes formed as the result of contraction is
\be \label{rhobh}
\rho_{BH} \sim \frac{M}{r_g^3} \sim \frac{\rho_1}{x^3}
\ee
where $x=\frac{r_g}{r_1}=2 \rho_1 G(t_{BH}) t_1^2$ with
 $r_g=2G(t_{BH})M$ is the gravitational radius of considered configuration having mass $M$.\\
So
\be \label{rhobhf}
\rho_{BH} \sim \frac{(t_{BH})^{3n}}{8 \rho_0^2 G_0^3 t_0^{4+n} t_1^{2+2n}} 
\ee
where $t_{BH}$ is the time at which PBH is formed.\\
The maximal density which may be reached in the contraction of non-spherical configuration is given by \cite{ky}
\be \label{maxr}
\rho_{max}=\frac{\rho_1}{S^3}  .
\ee
In order to form the black hole, the configuration should be nearly spherically symmetric. i.e.
\be \label{le}
S \leq x < 1  .
\ee
The upperbound $x<1$ gives
\be \label{tbh}
2G(t_{BH})M<r_1 .
\ee
But the formation mass of a PBH at a particular time must be some fraction ($\eta$) of the mass contained within the cosmological horizon at that time. i.e. $M=\eta G^{-1} t$. This equation in conjuction with equation (\ref{tbh}) leads to an expression for formation time of PBH as
\be \label{tbh2}
t_{BH}<\frac{t_1}{2\eta} . 
\ee
Thus the formation time of a PBH remains between $t_1$ and $\frac{t_1}{2\eta}$.\\
 i.e.
\be \label{tbh3}
t_1<t_{BH}<\frac{t_1}{2\eta} .
\ee
For matter-dominated era, $\eta$ is much less than 1 \cite{kp,khlopov,cgl} .
Thus, for a typical time of gravitational instability leading to formation of PBH ($t_1 \sim 10^{-34}$ s) and $\eta \sim 10^{-4}$, one gets $t_{BH} \sim 10^{-30}$ s. 

~~\\
The sufficient condition for the PBH formation imposes constraint on the 
inhomogeneity of the density distribution of the configuration at 
time $t_{BH}$ in the form \cite{ky}
\be \label{del}
\frac{\delta \rho_{BH}}{\rho_{BH}} < 1
\ee
which is also satisfied in our case where $\frac{\delta \rho_{BH}}{\rho_{BH}} =3n\frac{\delta t_{BH}}{t_{BH}}$ with $n \sim 0.00007$ and $\frac{\delta t_{BH}}{t_{BH}} < 1$.\\
We, thus, arrive at the conclusion that PBHs can indeed be formed in the matter-dominated era within Brans-Dicke formalism. \\

\section{PBH evaporation}
To study PBH evaporation, we consider an early matter-dominated era between the epochs of inflation and reheating. Keeping in mind that presence of early matter-domination should not affect the period of nucleosynthesis, we can extend reheating time upto $10^{-18}$s before which presently evaporating PBHs could be formed.\\ 
The rate at which the PBH mass decreases due to Hawking evaporation is given by
\be \label{mev1}
\dot{M}_{evap}=-4 \pi r_{BH}^2 a_H T_{BH}^4  .
\ee
Using the standard expressions for the black hole radius $r_{BH}=2GM$ and the Hawking temperature $T_{BH}=\frac{1}{8 \pi GM}$, one gets
\be \label{mev2}
\dot{M}_{evap}=- \frac{a_H}{256 \pi^3} \frac{1}{G^2 M^2} 
\ee
where $a_H$ is the Stefan-Boltzmann constant multiplied with number of degrees of freedom available for radiation.

PBH formation and evaporation could be greatly modified in variable-$G$ cosmologies, since many of their properties (eg. their radius and Hawking temperature)
depend explicitly on $G$. However, the nature of the modification depends upon the extent to which a PBH preserves the value of $G$ at its formation epoch 
rather than following the background cosmological value. Barrow \cite{jdb} first drew attention to this problem and introduced two possibilities : `` scenario A'' where $G$ has the same value everywhere at a given time, so that PBH evaporation is always determined by its current value ; ``scenario B'' where the local value of $G$ within the black hole is preserved implying gravitational memory so that the evaporation is determined by the value of $G$ when the PBH is formed. Here, in our study of PBH evaporation, we consider the above two cases.
\subsection{Scenario A}
In this scenario, $G$ has the same value everywhere at a given time.\\
If early matter-dominated era exists upto reheating time $t_2$, then the evaporation equation becomes
\ba \label{wint1}
\int_{M_i}^M M^2 \,dM=-\frac{\alpha}{t_0^{2n}}\Big[\int_{t_i}^{t_2} t^{2n} \,dt + \int_{t_2}^{t_e} t_e^{2n} \,dt + \int_{t_e}^{t} t^{2n} \,dt \Big]
\ea
where $M_i=10^{-4} \times G^{-1}t_i$ is the initial mass of PBH formed at time $t_i$ in matter-dominated era, $\alpha=\frac{a_H}{256 \pi^3} \frac{1}{G_0^2} \approx \frac{1}{G_0^2} $, $t_e$ is the time of matter-radiation equality and $G$ has the form $G=G_0 (\frac{t_0}{t_e})^n$ in radiation-dominated era. \\
The evaporating equation for PBHs formed in early radiation-dominated era has the form
\ba \label{wint2}
\int_{M_i}^M M^2 \, dM=-\frac{\alpha}{t_0^{2n}} \Big[\int_{t_i}^{t_e} t_e^{2n} \, dt + \int_{t_e}^{t} t^{2n} \, dt \Big]  .
\ea
where $M_i=G^{-1}t_i$ is the initial mass of PBH formed at time $t_i$ in radiation-dominated era \\
For different initial times ($t_i$), the numerical solutions of equations (\ref{wint1}) and (\ref{wint2}) are exhibited in the Table-I. To arrive at the numbers, we have used  $t_e\approx 10^{11}$s and $t_2\approx 10^{-18}$s. The last two rows of the table give the formation times of presently evaporating PBHs formed in early radiation-dominated era and in early matter-dominated era respectively.
\begin{table}
\begin{tabular}[c]{|c|c|c|c|c|}
\hline
$t_i$ & $(M_i)_0$ & $M_i$ & $\tau_0$ & $\tau$ \\
\hline
$10^{-23}$s & $10^{15}$g & $1.01\times 10^{11}$g & $3.34\times 10^{16}$s & $3.39\times 10^4$s\\
\hline
$10^{-22}$s & $10^{16}$g & $1.01\times 10^{12}$g & $3.34\times 10^{19}$s & $3.39\times 10^7$s\\
\hline
$10^{-21}$s & $10^{17}$g & $1.01\times 10^{13}$g & $3.35\times 10^{22}$s & $3.39\times 10^{10}$s\\
\hline
$10^{-20}$s & $10^{18}$g & $1.01\times 10^{14}$g & $3.35\times 10^{25}$s & $3.39\times 10^{13}$s\\
\hline
$10^{-19}$s & $10^{19}$g & $1.01\times 10^{15}$g & $3.36\times 10^{28}$s & $3.39\times 10^{16}$s\\
\hline
$2.36\times 10^{-23}$s & $2.36\times 10^{15}$g & $2.38\times 10^{11}$g & $4.42\times 10^{17}$s & $4.48\times 10^{5}$s\\
\hline
$2.35\times 10^{-19}$s & $2.35\times 10^{19}$g & $2.37\times 10^{15}$g & $4.37\times 10^{29}$s & $4.42\times 10^{17}$s\\
\hline
\end{tabular}
\caption{Evaporation times of PBHs having different formation times are shown in the table with symbols
$(M_i)_0 \sim$ Initial mass of PBH formed in early radiation-dominated era,
$M_i \sim$ Initial mass of PBH formed in early matter-dominated era,
$\tau_0 \sim$ Evaporation time of PBH formed in early radiation-dominated era   and
$\tau \sim$ Evaporation time of PBH formed in early matter-dominated era.}
\end{table}

It is clear from Table-I that the PBHs which are formed during early matter-dominated era will evaporate in significantly quicker rate than those of early radiation-dominated era because of their low masses at the time of formation.
To facilitate better comparison, we plot a graph between formation time $t_i$ and evaporating time $t_{evap}$ of PBHs for both early radiation-dominated era and early matter-dominated era in Figure-1.

\begin{figure}[h]
\includegraphics{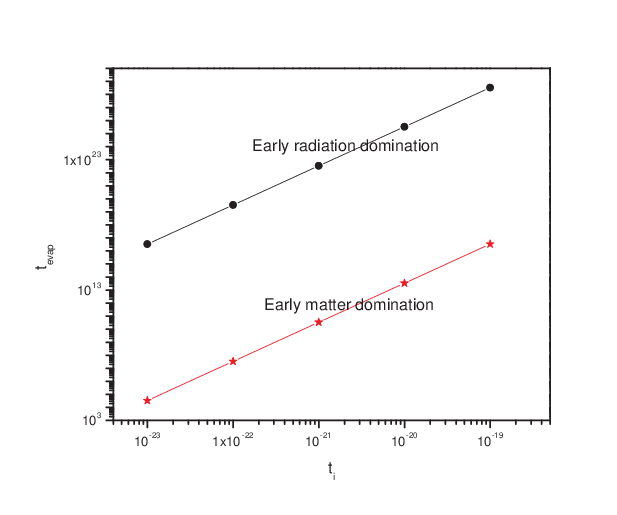}
\caption{Variation of evaporation times of PBHs $(t_{evap})$ with their formation times $(t_i)$ for both early radiation-dominated era and early matter-dominated era are shown in the Figure. For simpicity, here we used logarithmic scale in both axes.}
\label{fig1}
\end{figure}

\subsection{Scenario B}
In this scenario, $G$ associated with the black hole will continue to hold its value when PBH formation started.\\
Considering continuation of the  matter-dominated era upto reheating time $t_2$, the evaporation equation takes the form
\be \label{mint1}
\int_{M_i}^M M^2 \,dM=-\frac{\alpha}{t_0^{2n}} \int_{t_i}^{t} t^{2n} \,dt
\ee
where $M_i=10^{-4} \times G^{-1}t_i$ is the initial mass of PBH formed at time $t_i$ in matter-dominated era.\\
By integrating, we get
\be \label{mval1}
M^3=M_i^3+\frac{3\alpha}{t_0^{2n}} \frac{1}{2n+1} (t_i^{2n+1}-t^{2n+1})
\ee
which gives evaporation time of PBH as
\be \label{tau1}
\tau=\Big[t_i^{2n+1}+\frac{t_0^{2n}}{3\alpha} (2n+1) M_i^3 \Big]^{\frac{1}{2n+1}}
\ee
But for PBHs which are formed in early radiation-dominated era, the evaporating equation takes the form,
\be \label{mint2}
\int_{M_i}^M M^2 \, dM=-\alpha \Big(\frac{t_0}{t_e}\Big)^{2n} \int_{t_i}^{t_e}  \, dt
\ee
On integration, one gets
\be \label{mval2}
M^3=M_i^3+3\alpha\Big(\frac{t_0}{t_e}\Big)^{2n} (t_i-t)
\ee
where $M_i=G^{-1}t_i$ is the initial mass of PBH formed at time $t_i$ in radiation-dominated era. \\

Equation (\ref{mval2}) gives evaporation time of PBH as
\be \label{tau2}
\tau=t_i+\Big(\frac{t_0}{t_e}\Big)^{2n} \frac{1}{3\alpha} M_i^3 .
\ee
For different initial times ($t_i$), the results from equations (\ref{tau1}) and (\ref{tau2}) are shown in the Table-II.
\begin{table}
\begin{tabular}[c]{|c|c|c|c|c|}
\hline
$t_i$ & $(M_i)_0$ & $M_i$ & $\tau_0$ & $\tau$ \\
\hline
$10^{-23}$s & $10^{15}$g & $1.01\times 10^{11}$g & $3.34\times 10^{16}$s & $3.38\times 10^4$s\\
\hline
$10^{-22}$s & $10^{16}$g & $1.01\times 10^{12}$g & $3.34\times 10^{19}$s & $3.39\times 10^7$s\\
\hline
$10^{-21}$s & $10^{17}$g & $1.01\times 10^{13}$g & $3.34\times 10^{22}$s & $3.39\times 10^{10}$s\\
\hline
$10^{-20}$s & $10^{18}$g & $1.01\times 10^{14}$g & $3.34\times 10^{25}$s & $3.39\times 10^{13}$s\\
\hline
$10^{-19}$s & $10^{19}$g & $1.01\times 10^{15}$g & $3.34\times 10^{28}$s & $3.39\times 10^{16}$s\\
\hline
$2.37\times 10^{-23}$s & $2.37\times 10^{15}$g & $2.38\times 10^{11}$g & $4.42\times 10^{17}$s & $4.48\times 10^{5}$s\\
\hline
$2.35\times 10^{-19}$s & $2.35\times 10^{19}$g & $2.37\times 10^{15}$g & $4.35\times 10^{29}$s & $4.42\times 10^{17}$s\\
\hline
\end{tabular}
\caption{Evaporation times of PBHs for different formation times are shown in the table with symbols
$(M_i)_0 \sim$ Initial mass of PBH formed in early radiation-dominated era,
$M_i \sim$ Initial mass of PBH formed in early matter-dominated era,
$\tau_0 \sim$ Evaporation time of PBH formed in early radiation-dominated era   and
$\tau \sim$ Evaporation time of PBH formed in early matter-dominated era.}
\end{table}

Here, again, one finds that the PBHs which are formed during early matter-dominated era will evaporate in significantly quicker rate than those of early radiation-dominated era.

From Table-I and Table-II, one finds that the gravitational memory does not significantly affect the longevity of the PBHs.

In recent papers \cite{nsm,mgs}, it is shown that the accretion of radiation 
prolongates the lifetime of PBHs. Now one can consider the accretion of 
radiation for this case also. Here the initial mass of PBH is $10^{-4}$ times 
the horizon mass$(M_H)$ and PBH undergoes evaporation during early 
matter-dominated era. So during the radiation-dominated era, 
the initial mass of previously formed PBH becomes much smaller 
than $10^{-4}M_H$. 
But we know that the accretion of radiation is negligible if initial mass of 
PBH is less than $0.01M_H$ \cite{ns2}. Hence the accretion of radiation is 
ineffective for the PBHs which are formed in early matter-dominated era.  

\section{Constraints on PBH}
We now discuss the different cosmological constraints associated with PBHs 
formed in matter-dominated era.  PBH whose lifetime exceeds present era 
will contribute to the overall energy density.
As the present observable Universe is nearly flat and, therefore,
possesses critical density, the PBH mass density can be constrained
on the ground that it should not overdominate the Universe.
PBHs evaporate by producing  bursts of evaporation products.
Limits can be obtained by imposing that they should not interfere 
disastrously with esablished processes such as those of nucleosynthesis. 
Shorter-lived PBHs will have evaporated completely at an earlier stage. 
If this happened well before photon decoupling time, then their 
Hawking radiation will thermalize with the surroundings,
boosting the photon-to-baryon ratio \cite{za}. In the case of evaporation after
photon decoupling, the radiation spectrum is  affected and subsequently
redshifts in a monotonic manner. Thus, constraints arise from the cosmic
background radiation at high frequencies \cite{mc, ph, carr}. 
Further, if the PBHs evaporate
close to the time of photon decoupling, it cannot be fully thermalised
and will produce distortion in the cosmic microwave background spectrum.
Generally speaking, at a given epoch, the constraint on various physical
observables is usually dominated by those PBHs with a lifetime of order of
the epoch in question. Hence, the observational constraint can be translated
into an upper
limit on the initial mass fraction of PBHs.
In terms of mass fraction, the different astrophysical constraints 
associated with PBHs formed in the early radiation-dominated era have recently 
been analysed by Nayak, Majumdar and Singh \cite{nms}. In this work, 
we analyse the corresponding  
constraints for PBHs formed in the early matter-dominated era by using 
the result of our previous work \cite{nms} and conversion formula.
 
The fraction of the Universe's mass going into PBHs of mass M at a time $t$ 
in radiation-dominated era within BD formalism is given by
\be \label{beta1}
\beta_0(M)=\frac{\rho_{PBH,M}(t)}{\rho_{tot}(t)}
\ee

If there exists an early stage of matter-domination immediately after inflation, then the constraints on the fraction of the Universe going into PBHs during the matter-dominated era and $\beta_0(M)$ are related via the equation \cite{pk}
\be \label{beta2}
\beta(M)=\beta_0(M) \eta^{\frac{1}{2}} \Big(\frac{t_2}{t_{pl}}\Big)^{\frac{1}{2}} \Big(\frac{M}{M_{pl}}\Big)^{-\frac{1}{2}}
\ee
with $M=\eta (\frac{t}{t_{pl}}) M_{pl}$ .\\

But for early radiation-dominated era,  $\rho_{tot}(t) \approx \rho_{PBH}(t)+\rho_{rad}(t)$. So $\beta_0(M)$ becomes
\be \label{betaf}
\beta_0(M)=\frac{\alpha_0(M)}{1+\alpha_0(M)}
\ee
where $\alpha_0(M)$ is the initial mass fraction of PBH defined as
\be \label{alpha}
\alpha_0(M)=\frac{\rho_{PBH,M}(t)}{\rho_{rad}(t)}.
\ee
So for $\alpha_0(M) << 1$ as found in our earlier work \cite{nms}, $\beta_0(M) \approx \alpha_0(M)$ .

\begin{table}
\begin{tabular}[c]{|c|c|c|c|}
\hline
Cause of the constraint & $M$ in gm  & $\beta_0(M)<$ & $\beta(M)<$\\
\hline
Present Density & $2.36 \times 10^{15}$ & $3.43\times 10^{-18}$ & $7.04\times 10^{-18}$\\
\hline
Photon Spectrum & $2.36 \times 10^{15}$ & $5.34 \times 10^{-26}$ & $1.09\times 10^{-25}$\\
\hline
Distortion of CMB & $1.74 \times 10^{12}$ & $1.28\times 10^{-21}$ & $9.66\times 10^{-20}$\\
\hline
Helium abundance & $2.28 \times 10^{10}$ & $3.82\times 10^{-19}$ & $2.53\times 10^{-16}$\\
\hline
Deuterium abundance & $2.28 \times 10^{10}$ & $5.10\times 10^{-21}$ & $3.38\times 10^{-18}$\\
\hline
\end{tabular}
\caption{The values of differeent constraints associated with PBHs formed in early radiation-dominated era and in early matter-dominated era are presented in the Table in separate columns.}
\end{table}

We can obtain the bounds on $\beta(M)$ using the translation formula 
(\ref{beta2}) and the constraints we have obtained in our previous work 
\cite{nms}. We have presented the constraints for both early radiation-dominated era and early matter-dominated era in Table-III for easy 
comparison. Here we have used $\eta=10^{-4} \times \Big(\frac{t}{t_0}\Big)^n$ 
which comes from our assumptions of early matter domination i.e. 
$M_i=10^{-4} \times G^{-1}t_i$ and Brans-Dicke gravity i.e. $G=G_0 \Big(\frac{t_0}{t}\Big)^n$.
Here one finds a significant change in the value of different constraints from 
their early radiation-dominated era counterparts. 
All known constraints associated with evaporation of PBHs 
tend towards the larger value than those of early radiation-dominated era. Nucleosynthesis constraints 
(Helium abundance constraint and Deuterium abundance constraint) are 
larger by three orders in magnitude whereas constraints associated with 
presently evaporating PBHs are larger by less than one order in magnitude.

It may be mentioned here that the recent
paper by Carr et al. \cite{carretal} has reanalyzed
the standard constraints on primordial black holes by considering the effects
of emission of quarks and gluons and the resultant secondary emission of
photons. It was shown that the effect of secondary photon emission could
alter the standard constraints on PBH fraction by a couple of orders of
magnitude in certain cases, e.g., deuterium constraint, while leaving
the standard constraints more or less unaltered in certain other cases,
e.g., distortion of CMB spectrum. The results obtained by Carr et al.
\cite{carretal} are
based on detailed numerical analysis. Of course, such a scenario of
emission would also
impact constraints on Brans-Dicke primordial black holes in more or less
similar ways as they impact PBHs in the standard cosmology. However, the
lack of analytical results describing the effect of such emission on
the constraint formalism makes it considerably harder to perform  a similar
analysis in the context of an altered gravitational scenario. 
So following the arguments of our previous paper \cite{nms}, here we can only write that the emission of quarks and gluons and the resultant secondary emission of photons may affect the constraints associated with matter-dominated era PBHs 
in similar fashion as they affect their radiation-dominated era counterparts.
\section {Conclusions}
We have shown that PBHs can be formed in matter-dominated era of cosmic
evolution with gravity described by BD theory. For realisation of our result,
we consider an early matter-dominaed era existing between end of inflation and begining of reheating.
We found that the evaporation period of those PBHs are substantially 
shorter than those of the early
radiation-dominated era. Thus, in comparison with latter case, less number of PBHs could exist today. Further, we got that the constraints on PBH formation 
tend towards the larger value in the present formalism
compared with their radiation-dominated counterparts, which provides a strong evidence for
the enhancement of PBH formation in matter-dominated era than the
radiation-dominated era. It may be recalled that in matter-dominated phase,
cosmic matter remains in a nearly zero pressure state. Absence of an opposing
force to gravitational pull, thus, increases the probability of PBH formation
in this era. Our work presents yet another application of BD theory as a viable alternate
theory of gravity in addition to the various
ones mentioned in the introduction.

\section*{Acknowledgements}
We are thankful to Institute of Physics, Bhubaneswar, India, for providing the library and computational facility. B.Nayak would like to thank the Council of Scientific and Industrial Research, Government of India, for the award of SRF, F.No. $09/173(0125)/2007-EMR-I$ .


\end{document}